# Stacking-order-dependent electronic properties of $MoTe_2/WSe_2$ moiré bilayers

Zhongdong Han[1,2,3], Wenjin Zhao[4], Eegene Clara Chung[2,5], Chia-Hao Lee[5], Zui Tao[5], Zhengchao Xia[5], Yichi Zhang[2], Yiyu Xia[5], Jekwan Lee[1,2,3], Bowen Shen[5], Ariana Ray[2], Yu-Tsun Shao[5], Tingxin Li[5], Shengwei Jiang[1,2], Yihang Zeng[1,2], Kenji Watanabe[6], Takashi Taniguchi[6], David Muller[4,5], Kin Fai Mak[1,2,3,4,5†], Jie Shan[1,2,3,4,5†]

[1]Laboratory of Atomic and Solid State Physics, Cornell University, Ithaca, NY, USA
[2]Department of Physics, Cornell University, Ithaca, NY, USA
[3]Max Planck Institute for the Structure and Dynamics of Matter, Hamburg, Germany
[4]Kavli Institute at Cornell for Nanoscale Science, Ithaca, NY, USA
[5]School of Applied and Engineering Physics, Cornell University, Ithaca, NY, USA
[6]National Institute for Materials Science, Tsukuba, Japan

[†]Email: kin-fai.mak@mpsd.mpg.de; jie.shan@mpsd.mpg.de

**Abstract**
Transition metal dichalcogenide (TMD) moiré bilayers have realized a wide range of strongly correlated and topological phenomena. The physics in these materials is often sensitive to the interlayer stacking order. Polarization-resolved optical second harmonic generation (SHG) is the most used technique for stacking order characterization but unverified for most heterobilayers. Here we calibrate the optical SHG for angle-aligned $MoTe_2/WSe_2$ bilayers by the scanning transmission electron microscopy (STEM). We directly compare the transport and magnetic properties and the electronic phase diagram for two distinct stacking orders. With the calibrated stacking order assignment, we clarify the interpretation of earlier results, including the nature of the Chern insulator, mechanism of an electric-field-tuned metal-insulator transition at half band filling, and the Kondo lattice physics. Our work provides a consistent picture of the relation between the stacking order and the electronic properties of $MoTe_2/WSe_2$ moiré bilayers.

**Main**
Semiconductor moiré materials have emerged as a highly tunable platform for studying electronic correlations and topology. A wide range of phenomena, including Mott and Wigner-Mott insulators[1-14], kinetic antiferromagnetism and spin polarons[15-19], Kondo lattice physics[20-28], and the quantum anomalous Hall[29-36] and quantum spin Hall effects[29,34,37,38], has been demonstrated in one system--angle-aligned $MoTe_2/WSe_2$ vertical heterobilayers[39-42]. Many of these phenomena are sensitive to stacking order, that is, whether the two transition metal dichalcogenide (TMD) monolayers are parallel (AA-stacking) or antiparallel (AB-stacking) to each other (Fig. 1a,b). Angle-aligned $MoTe_2/WSe_2$ bilayers form a moiré lattice with period $a_{\mathrm{M}} \approx 5$ nm because of the 7% lattice mismatch between the two TMDs. The electronic states of interest are originated from the valence band in the K/K' valley of monolayer $MoTe_2$ or $WSe_2$, where the presence of a strong Ising spin-orbit interaction creates a large spin splitting in each valley and the spin and valley degrees of freedom are locked[41,43]. Because the moiré momentum is negligible compared to the atomic scale momentum, moiré bands can hybridize only with those from the same valley (Fig. 1c). The spin-valley locking renders the moiré band structure stacking dependent: interlayer hybridization is spin allowed for AA-stacking and spin forbidden for AB-stacking[31]. This stacking-dependent interlayer hybridization, however, poses significant challenges in understanding several earlier experimental results, including the nature of the Chern insulator, mechanism of an electric-field-tuned metal-insulator transition at half band filling, and the Kondo lattice physics[1,20,29,37].

In all previous experiments, stacking order in angle-aligned $MoTe_2/WSe_2$ bilayers was determined by the polarization-resolved optical second harmonic generation (SHG)[44,45]. Specifically, monolayer TMDs lacking inversion symmetry allow for a second-order response; the second harmonic (SH) dipole moment is uniquely determined by the crystal's symmetry[46-48]. In angle-aligned bilayers, the SH dipole from each monolayer is expected to be in-phase (out-of-phase) for AA-(AB-)stacking, leading to constructive (destructive) interference of the SH radiation. This convenient optical technique has been extensively studied in natural TMD bilayers and multilayers[46-48] but remained unverified for heterobilayers except $WS_2/WSe_2$ (ref.[44,45]). In this work, we directly determine the stacking order in angle-aligned $MoTe_2/WSe_2$ bilayers by the scanning transmission electron microscopy (STEM) and establish correlation between the stacking order and SHG. Contrary to homobilayers, we observe destructive interference in SHG for AA-stacked $MoTe_2/WSe_2$ bilayers with excitation at 800 nm. Based on the revised stacking order assignment, we present a consistent picture of stacking-dependent electronic phase diagrams of the material.

**Structure characterization**

We directly determine the structure of $MoTe_2/WSe_2$ bilayers by atomic-resolution imaging using STEM. The annular dark-field (ADF) image intensity scales approximately with $Z^{1.7}$, where $Z$ is the atomic number (Ref.[49]), and the contributions from overlapping chalcogen atoms are summed (Table 1). The atoms in each TMD monolayer can be clearly identified: the W sites appear about two times brighter than the $Se_2$ sites in $WSe_2$, and the $Te_2$ sites are about three times brighter than the Mo sites in $MoTe_2$. But the W and $Te_2$ sites have similar intensities, making accurate identification of the atoms and consequently stacking order in heterobilayers difficult.

We designed a special sample geometry to unambiguously determine the stacking order in $MoTe_2/WSe_2$ bilayers by STEM and correlate it with SHG (see Methods for details on the sample fabrication and STEM and SHG measurements). Figure 2a and 2b show two $MoTe_2/WSe_2$ samples angle aligned based on SHG. The $MoTe_2$ and $WSe_2$ monolayers in these samples were cut from the same flakes and one $MoTe_2$ layer was rotated by 180° relative to the other. The two bilayer samples are thus guaranteed to have different stacking orders. Figure 2d and 2e are the ADF-STEM images of the isolated $WSe_2$ and $MoTe_2$ monolayer regions of sample 1 (Fig. 2a). Intensity profile analysis of these images (Fig. S1) clearly identifies the metal and chalcogen atoms. We determine the overlapped region to be AB-stacked based on the relative orientations of the metal sub-lattices (denoted by yellow dashed lines) in the isolated monolayer regions. Similarly, we determine sample 2 (Fig. 2b) to be AA-stacked based on the ADF-STEM images in Fig. 2g,h (see Fig. S2,3 for STEM images of the bilayer regions for further confirmation of the stacking order assignment).

We correlate the SHG with stacking order by performing SH measurements on both the isolated monolayer regions and the bilayer region (Fig. 2c,f). We used the reflection geometry with normal incidence for excitation at 800 nm. The SH intensity in cross polarization exhibits a six-fold symmetric response as a function of the linear polarization angle of the excitation beam, or equivalently, the crystal's azimuthal angle. Such a response is consistent with the three-fold rotational symmetry of monolayer TMDs[46-48] and agrees with previous reports[1,29]. The SH intensity is comparable for monolayer $MoTe_2$ and $WSe_2$. But surprisingly, it nearly vanishes for the AA-stacked sample (Fig. 2f) and is about 2-3 times of that of each monolayer for the AB-stacked sample. The result highlights that the SH response is a frequency-dependent complex function and can be significantly altered when either the fundamental or SH radiation is close to an electronic transition. Given the large difference in two TMDs' band gaps and the 800-nm excitation lying between the two gap values, a phase shift between the SH dipole moments from two layers in AA-stacked bilayers is generally expected. This is further supported by the excitation wavelength dependence of SHG from another AA-stacked $MoTe_2/WSe_2$ bilayer (Fig. S4). We observe destructive interference of the SH radiation for all three excitation wavelengths (800 nm, 1100 nm and 1220 nm), with most complete cancelation at 800 nm.

**Electronic phase diagrams**

Equipped with the calibrated SHG, we study correlation between the stacking order and the electronic phase diagram in dual-gated $MoTe_2/WSe_2$ devices by combining the electrical transport and magneto optical dichroism (MCD) measurements. The latter characterizes the magnetic properties of the bilayers. (See Methods and Fig. S5 for stacking-order-dependent exciton hybridization). The two gates allow independent control of the lattice filling factor ($\nu$) of holes in the bilayer and the electric field ($E$) perpendicular to the sample plane. Compared to earlier works, we used thinner hexagonal boron nitride (hBN) gate dielectrics to access higher breakdown fields and wider phase diagrams[22]. Unless otherwise specified, the sample temperature was 1.6 K. See Methods for details on device fabrication and electrical and magneto-optical measurements.

The results for an AB-stacked $MoTe_2/WSe_2$ bilayer are shown in the left column of Fig. 3. Figure 3a and 3b are the longitudinal ($R_{xx}$) and Hall ($R_{xy}$) resistances, respectively, as a function of $\nu$ and $E$ under a perpendicular magnetic field $B = 14$ T. The resistance maps are dominated by stripes, originating from the Shubnikov-de Haas (SdH) oscillations in the $WSe_2$ layer, which has a more dispersive moiré valence band and higher hole mobility. (Substantially weaker SdH oscillations in the $MoTe_2$ layer can also be discerned especially in the region of $1 < \nu < 2$). Diagonal stripes show that both layers are charge compressible and the partition in doping is tunable by electric field. In contrast, vertical stripes show that the $MoTe_2$ layer is incompressible and the hole density in the $WSe_2$ layer is independent of electric field.

We construct the electrostatic phase diagram in Fig. 3e from the transport data. In the yellow region ($\nu_{Mo} + 0$, left to the red solid line), only the $MoTe_2$ layer is doped and the $WSe_2$ layer is charge neutral ($MoTe_2/WSe_2$ is a type I heterojunction at $E = 0$ with the $MoTe_2$ band gap lying entirely inside the $WSe_2$ band gap). Here $\nu_{Mo}$ and $\nu_W$ denote the hole filling factors of the $MoTe_2$ and $WSe_2$ layers, respectively. The resistance in this region is generally high, likely due to the flat $MoTe_2$ Hubbard bands, which favor charge localization. The other limit is the high $E$-field region ($0 + \nu_W$), where the $WSe_2$ layer is doped and the $MoTe_2$ layer is charge neutral. We could not access this region experimentally due to dielectric breakdown. In the two middle blue regions, the stripes are vertical, showing that the Fermi level is inside a gap of the $MoTe_2$ layer—a Mott gap for ($1 + \nu_W$) and a moiré band gap for ($2 + \nu_W$)—and only the $WSe_2$ layer hosts itinerant holes. The rest phase regions bear generic filling factors.

The constructed phase diagram is supported by the magnetic response of the bilayer. Figure 3c and 3d are the spectrally averaged MCD over the attractive polaron resonance of $MoTe_2$ as a function of $\nu$ and $E$ at $B$ = 1 T and 8.9 T, respectively. Strong MCD response is observed mainly around $\nu_{\text{Mo}} = 1$ (denoted by red dashed lines) and in region $(1 + \nu_{\text{W}}$, marked by black dashed lines). This is consistent with the $MoTe_2$ layer being a Mott insulator at half band filling; the spins after charge localization (i.e. local magnetic moments) are highly polarizable, giving a strong MCD response[44] relative to the other phase regions. A large MCD response is also observed in a region above $(1 + \nu_{\text{W}})$ at 1 T. This presumably arises from the $MoTe_2$ band edge, which shifts out of the experimental window at 8.9 T (Fig. 3d).

The corresponding results for an AA-stacked $MoTe_2$/$WSe_2$ bilayer are shown side by side in Fig. 3 (see Fig. S6,7 for comparison of resistance and MCD at $B$ = 0). They are consistent with previous reports[20,29,37]. The resistance maps (Fig. 3f,g) share some similarities with that of the AB-stacked sample but are substantially richer. Similarly, we observe vertical resistance stripes in region $(0 + \nu_{\text{W}})$, $(1 + \nu_{\text{W}})$ and $(2 + \nu_{\text{W}})$, and field-dispersive stripes in region $(1 < \nu_{\text{Mo}} < 2)$. On the other hand, we find vertical instead of diagonal stripes in region $(0 < \nu_{\text{Mo}} < 1)$ and new insulating states at $\nu = 1$ and $\nu = 2$. Particularly, the insulating state at $\nu = 1$ for the intermediate $E$-fields, exhibiting vanishing $R_{\text{xx}}$ and quantized $R_{\text{xy}}$, is a Chern insulator (Fig. 3j). In the presence of finite interlayer hybridization, the layer-resolved filling factors ($\nu_{\text{Mo}}$ and $\nu_{\text{W}}$) become approximate.

The MCD maps under $B$ = 1 T (Fig. 3h) and 8.9 T (Fig. 3i) also show some similarities and differences between the two stacking orders. Specifically, a strong MCD response is also observed around $\nu_{\text{Mo}} = 1$ and in region $(1 + \nu_{\text{W}})$ for the AA-stacked sample, where the $MoTe_2$ layer is a Mott insulator. However, the strong MCD response in region $(1 + \nu_{\text{W}})$ terminates at an arc-shaped boundary, which expands into higher fillings with increasing magnetic field, whereas for the AB-stacked sample, strong MCD persists throughout the entire region. In addition, strong MCD is observed around $\nu = 1$ for the intermediate $E$-fields, in accord with the Chern insulator.

The results in Fig. 3 are consistent with the stacking dependent interlayer hybridization. The most important difference between the two stacking orders is how the SdH oscillations behave in regions with generic fillings. The AB-stacked samples have negligible interlayer hybridization. Therefore, the Landau levels in two layers are decoupled; both can be populated by gating and shifted in energy by electric field; the SdH stripes are diagonal in regions with generic fillings. Conversely, the AA-stacked samples have finite interlayer hybridization, weakening the electric-field effects. The Landau levels are layer hybridized in region $(0 < \nu_{\text{Mo}} < 1)$ and the stripes are field independent. The stripes show field dispersion in region $(1 < \nu_{\text{Mo}} < 2)$ but are still substantially different from the layer-decoupled case, indicating weaker but non-negligible interlayer hybridization. Furthermore, the mixed valence topological insulator (MVTI) exhibiting the quantum spin Hall effect at $\nu = 2$ and the Chern insulator at $\nu = 1$ (in the presence of strong Coulomb interactions[31-35,50,51]) require interlayer hybridization and are observed only in AA-stacked samples (Fig. 3j).

**Revised interpretation of earlier results**

With better understanding of the relationship between the stacking order and electronic phase diagram, we clarify the interpretation of several earlier studies. The first concerns the nature of the Chern insulator at $\nu = 1$. An earlier study assigned the samples as AB-stacked bilayers based on the uncalibrated SHG[29]. By studying the layer-resolved MCD response, Ref.[52] further concluded that the Chern state is valley coherent. Based on the corrected stacking order assignment and the reported layer-resolved MCD results[52], we conclude that the Chern insulator is a valley-polarized state, which is consistent with most theoretical expectations[31-35].

The second is the metal-insulator transition induced by electric field at fixed filling $\nu = 1$ (Ref.[1]). The transition was thought to occur in AA-stacked samples based on the uncalibrated SHG. The metal-insulator transition was interpreted as a bandwidth-tuned Mott transition because electric field in AA-stacked bilayers can tune the moiré bandwidth like in small-angle twisted homobilayers[50,51,53-64]. With the revised stacking order assignment, we note that moiré bandwidth tuning by electric field is unlikely due to the negligible interlayer hybridization in AB-stacked bilayers. But electric field can tune the charge transfer gap between the $WSe_2$ moiré band and the $MoTe_2$ Hubbard band through the linear interlayer Stark effect. The charge transfer gap, acting as a Mott gap, is continuously tuned to zero at the metal-insulator transition; such a transition belongs to the same universality class as the bandwidth-tuned Mott transition[65].

The last concerns the Kondo lattice phenomena observed in region $(1 + \nu_W)$ of the phase diagram. The samples were assigned as AB-stacked bilayers based on the uncalibrated SHG previously[20], but their negligible interlayer hybridization would imply a negligible Kondo coupling. The difficulty is resolved by the revised stacking order assignment: the finite interlayer hybridization in AA-stacked samples produces a measurable Kondo coupling between the local moments in the $MoTe_2$ layer and the itinerant holes in the $WSe_2$ layer. The arc-shaped boundary in the MCD response under a fixed $B$-field (Fig. 3h,i) separates a light Fermi liquid and a heavy Fermi liquid as reported in Ref.[66]. To the left of the boundary, the Zeeman field overcomes the Kondo coupling, the local moments and the itinerant holes are decoupled, and the polarized local moments give a strong MCD response. To the right of the boundary, the local moments and the itinerant holes are coupled to form Kondo singlets, resulting in a weak MCD response. The shape of the boundary is given by the dependence of Kondo coupling on the filling factor and band alignment (or electric field). In contrast, the strong MCD or local moment response is observed throughout region $(1 + \nu_W)$ for AB-stacked bilayers (Fig. 3c,d).

In summary, stacking order plays a crucial role in the electronic properties of TMD moiré bilayers because interlayer hybridization is stacking-order dependent. In this study, we calibrated the optical SHG—a commonly used technique for the stacking order characterization—by atomic resolution imaging using the STEM. We directly compared the transport and magnetic properties and the electronic phase diagram of AA- and AB-stacked $MoTe_2/WSe_2$ bilayers. With the revised stacking order assignment, we resolved difficulties in the interpretation of earlier results, including the nature of the Chern insulator, mechanism of an electric-field-tuned metal-insulator transition at half band filling, and the Kondo lattice physics.

## Methods

### STEM sample fabrication

Samples for the scanning transmission electron microscopy (STEM) measurements were fabricated using the 'tear and stack' method[67]. In short, $MoTe_2$ and $WSe_2$ monolayers were exfoliated from bulk crystals onto 285-nm $SiO_2$/Si substrates and identified by their reflection contrast. Polarization-resolved optical second harmonic generation (SHG) was performed on each monolayer; the second harmonic (SH) intensity exhibits a six-fold rotational symmetry, revealing the TMD crystallographic orientation. Each TMD monolayer was then cut into two halves using an AFM (atomic force microscopy) tip and two $MoTe_2/WSe_2$ bilayers with distinct stacking orders were assembled using one half from each flake. In both samples, the two TMDs were angle aligned according to their SHG, resulting in either parallel (AA) or antiparallel (AB) stacking order. A 180° rotation was introduced in the two halves of $MoTe_2$.

The heterostructures were assembled using the dry transfer method. First, a thin (3-5 nm) hexagonal boron nitride (hBN) flake was picked up by a polycarbonate (PC) film on a polydimethylsiloxane (PDMS) stamp. Next, the TMD flakes were picked up sequentially with the desired angle alignment. Last, the sample was covered by a thin ( ~ 1 nm) graphite capping layer to prevent oxidation of $MoTe_2$. All processes involving $MoTe_2$ prior to encapsulation were carried out in a nitrogen environment to minimize degradation.

Each sample has three distinct regions: two isolated monolayer regions and a bilayer region (Fig. 2a,b). Polarization-resolved SHG was performed on each region of the sample on the stamp. It was then released onto a TEM grid (PELCO 21589-10, silicon nitride membranes with holes). Annular dark-field STEM measurements were carried out on the same three regions to determine the lattice structure of the monolayers and their stacking order in the bilayer. Such a sample design enabled a direct correlation of SHG with the stacking order.

**Hall bar device fabrication**

Dual-gated Hall bar devices of angle-aligned $MoTe_2/WSe_2$ bilayers for combined electrical transport and optical measurements were fabricated using the method reported in previous studies[1,29]. Thin flakes required for device assembly—including few-layer graphite and hBN as gate electrodes and dielectrics, as well as monolayer TMDs—were exfoliated from bulk crystals onto 285-nm $SiO_2$/Si substrates and identified by their reflection contrast. Polarization-resolved SHG (calibrated by STEM above) was used to determine the crystallographic orientation of the TMD monolayers, angle align them and determine the stacking order of the bilayers.

A two-step stacking process was used to fabricate high-quality devices. First, hBN and graphite flakes were picked up and released onto a Si substrate with prepatterned electrodes to form the bottom gate. Thin (8 nm) platinum electrodes were deposited onto the hBN to define a Hall bar geometry using the standard electron-beam lithography and evaporation techniques. Next, the polymer residues on the device were removed by the AFM in the contact mode with a typical force of 500 nN. Last, the rest of the device, consisting of a graphite/hBN top gate and an angle-aligned $MoTe_2/WSe_2$ bilayer, was released onto the bottom-gate substrate. All processes involving $MoTe_2$ prior to encapsulation were carried out in a nitrogen environment to minimize oxidation. In these devices, the bottom gate fully covers the top gate, thereby eliminating parallel conduction channels, and the top-gate hBN dielectric[22] is substantially thinner (< 4 nm) than that in typical devices. Such a design enabled the application of high (up to 1.4 V/nm) electric fields perpendicular to the sample plane.

**STEM measurements**
STEM images were collected on two Thermo Fisher Spectra 300 aberration-corrected microscopes with an ultrahigh brightness cold field emission gun (X-CFEG). The acceleration voltage was 120 kV to minimize beam-induced damage. The probe-semi-convergence angle was 30 mrad with the probe current between 35-55 pA. The dwell times were 0.5 μs for 10-frame imaging and 5-10 μs for single frame imaging. The multi-frame images were rigid registered to mitigate scan distortions. Annular dark-field (ADF) images were acquired with inner and outer collection angles of 48 to 200 mrad to balance signal intensity with atomic Z-contrast. The pixel sizes ranged from 5.7 to 21 pm. The measurements were performed with samples at room temperature.

**Second harmonic measurements**
The optical SH measurements were performed in the reflection geometry using normal incidence excitation at room temperature. The output of a mode-locked Ti:sapphire oscillator (Spectra Physics, Tsunami) operating at an 80 MHz repetition rate was used as the excitation source. The pulses were of 100 fs duration and centered at 800 nm. The excitation beam passed a linear polarizer and an achromatic half-wave plate and was focused onto a spot size of about 1 μm on the sample by an objective with numerical aperture 0.6. The retroreflected SH signal (400 nm) was collected by the same objective, passed the same half-wave plate, and was separated by a beam splitter and filtered by a short-pass filter. An analyzer was used to select the polarization component that is perpendicular to the polarization of the excitation beam. The SH radiation was detected by an EMCCD (electron-multiplying charge-coupled device) camera (ANDOR iXon$^{EM}$+). The SH character was verified by its quadratic intensity dependence on the excitation intensity. The orientational dependence of the SH response was measured by rotating the achromatic half-wave plate which is equivalent to rotating the sample.

Polarization-resolved SHG has been employed to determine the crystal symmetry and crystallographic orientation of TMDs[46-48]. Monolayer TMDs ($D_{3h}$ symmetry) lacking inversion symmetry allow for a second-order response. The second-order susceptibility tensor $\chi^{(2)}$ contains one independent nonvanishing element, $\chi^{(2)}_{\mathrm{xxx}} = -\chi^{(2)}_{\mathrm{xyy}} = -\chi^{(2)}_{\mathrm{yxy}} = -\chi^{(2)}_{\mathrm{yyx}}$, where x and y axes correspond to the armchair and zigzag directions of the crystal, respectively. Excitation radiation with electric-field amplitude $E$ at the fundamental frequency induces a second-order nonlinear polarization $P^{(2)} \sim \chi^{(2)} : EE$, which contains components at the SH frequency. The SH intensity in cross polarization is expected to exhibit a six-fold symmetric response as a function of the crystal's azimuthal angle (or the linear polarization angle of the excitation beam), where the maxima (minima) correspond to the zigzag (armchair) directions of the crystal. In angle-aligned bilayers, each layer generates an SH response which adds constructively or destructively, depending on the stacking order. This effect is particularly pronounced in natural bilayers (with AB stacking order), where inversion symmetry is restored and the SH response vanishes. The method has not been verified for generic TMD heterobilayers.

**Electrical measurements**
Electrical transport measurements were performed in a closed cycle $^4$He cryostat (Oxford TeslatronPT) with magnetic fields up to 14 T and temperatures down to 1.5 K. Four-terminal resistance ($R_{\mathrm{xx}}$) and Hall resistance ($R_{\mathrm{xy}}$) were measured using the standard low-frequency lock-in techniques at 13.77 Hz with 1 mV AC excitation applied to the source electrode. Voltage drop between the probe electrodes and source-drain current were simultaneously recorded. All data were taken at 1.6 K unless otherwise specified.

**Magneto-optical measurements and analyses**

Reflective magneto circular dichroism (MCD) using normal incidence excitation was used to study the magnetic properties of the samples. Measurements were performed in a closed cycle $^4$He cryostat (attocube attoDry 2100) with magnetic fields up to 9 T and temperatures down to 1.6 K. A superluminescent light emitting diode (SLED) with broadband emission centered at 1050 nm was used as the excitation source. The SLED output was coupled into a single-mode fiber, collimated using a 10× objective, and focused onto a spot size of 1 μm on the sample using a low-temperature objective with numerical aperture 0.8. The incident power on the sample was kept below 30 nW to minimize the heating effects. The reflected light was collected by the same objective and detected by a spectrometer coupled with a liquid-nitrogen-cooled CCD (charge-coupled device). A combination of a linear polarizer and an achromatic quarter-wave plate was used to generate circularly polarized light. The MCD spectrum, $(R^+ - R^-)/(R^+ + R^-)$, was obtained by comparing the reflection spectrum $R^+$ and $R^-$ for left and right circularly polarized light, respectively. The spectrally averaged MCD over the attractive polaron resonance of $MoTe_2$ is illustrated in Fig. 3. It was obtained by averaging the absolute value of MCD over a wavelength range of 1068-1128 nm for AA-stacked samples and 1050-1120 nm for AB-stacked samples. All data were taken at 1.6 K unless otherwise specified.

**Stacking-dependent exciton hybridization**

Intralayer and interlayer optical transitions in $MoTe_2/WSe_2$ bilayers that share the same final (or initial) states can hybridize when the interlayer resonance is tuned by electric field to cross the intralayer resonance. Supplementary Fig. S5 illustrates the reflection contrast (RC) spectrum of $MoTe_2/WSe_2$ bilayers near the $WSe_2$ fundamental exciton resonance as a function of electric field at fixed filling factor $\nu = 1$. The RC spectrum, $(R - R_0)/R_0$, was obtained by comparing the raw reflection spectrum $R$ to a reference spectrum $R_0$ (which was measured at a high doping density with quenched exciton resonances). Hybridization between the intralayer and interlayer excitons is shown as level repulsion. The slope of the exciton energy as a function of the electric field (which is approximately the shift per unit electric field of the interlayer potential difference) corresponds to dipole moment ~ 0.28 e · nm. The value is consistent with $\frac{\varepsilon_{hBN}}{\varepsilon_{TMD}} et$, where $\varepsilon_{hBN}$ ~ 3 and $\varepsilon_{TMD}$ ~ 7 are the out-of-plane dielectric constants of hBN and TMD, respectively, and $t$ ~ 0.7 nm is the interlayer distance. Exciton hybridization is stacking-order dependent, occurring at positive fields (~ 0.3 V/nm) for AA-stacked samples and negative fields (~ -0.3 V/nm) for AB-stacked samples (a positive electric field shifts the energy bands of $WSe_2$ up relative to that of $MoTe_2$). They correspond to an interlayer potential difference of about 100 meV and -100 meV for the AA- and AB-stacked samples at $E$ = 0.

The stacking-dependent exciton hybridization in $MoTe_2/WSe_2$ can be understood considering the schematic energy band diagram (Supplementary Fig. S5a,b). The bilayers are a type I heterojunction with the $MoTe_2$ band gap lying entirely inside the $WSe_2$ band gap. The electronic states near the band edges are the K-valley states, which are split by the Ising spin-orbit coupling. In the AA-stacked samples, the spin-up valence bands are above the spin-down bands in both TMDs, whereas in the AB-stacked samples, the order of the two spin-split bands is reversed in $MoTe_2$. Since only the spin-conserved optical transitions (marked by arrowed lines) are dipole allowed, level repulsion is expected at a higher *E*-field in AA-stacked samples. This is fully consistent with experiment. The values of the *E*-fields for hybridization are dependent on the band offset and the difference in the intralayer and interlayer exciton binding energies. Given the spin splitting of ~ 100 meV for the valence bands in $MoTe_2$, the intralayer exciton binding energy is about 100 meV larger than the interlayer exciton binding energy.

**Acknowledgements**
This work was supported by the Gordon and Betty Moore Foundation (grant DOI 10.37807/GBMF11563; transport measurements) and the Air Force Office of Scientific Research under award number FA9550-20-1-0219 (optical measurements). E.C.C is supported by the Air Force Office of Scientific Research under award number FA9550-23-1-0688. C.-H.L. acknowledges the support by the Eric and Wendy Schmidt AI in Science Postdoctoral Fellowship, a program of Schmidt Sciences, LLC. Sample fabrication was carried out at the Cornell NanoScale Facility, an NNCI member supported by NSF grant NNCI-2025233. This work also used facilities of the Platform for the Accelerated Realization, Analysis, and Discovery of Interface Materials (PARADIM) and Cornell Center for Materials Research (CCMR) supported by NSF DMR-2039380 and DMR-1719875. The growth of the hBN crystals was supported by the Elemental Strategy Initiative of MEXT, Japan, and CREST (JPMJCR15F3), JST.

**Author contributions**
Z.H., W.Z. and Y.Z. fabricated the STEM samples. Z.H., Y.Z. and Y.X. fabricated the transport and optical devices. E.C.C. and C.-H.L. acquired and analyzed the presented ADF-STEM data, while A.R. and Y.-T.S. performed early STEM explorations under the supervision of D.A.M. Z.H., Z.T. and Z.X. performed the optical measurements. Z.H., W.Z., Z.X. and J.L. performed the SHG measurements. Z.H., Y.X. and Y.Z. performed the transport measurements. B.S., T.L., S.J. and Z. Y. performed early experiments under the supervision of K.F.M. and J.S. K.W. and T.T. grew the bulk hBN crystals. K.F.M. and J.S. designed the scientific objectives and oversaw the project. All authors discussed the results and commented on the paper.

## Figures

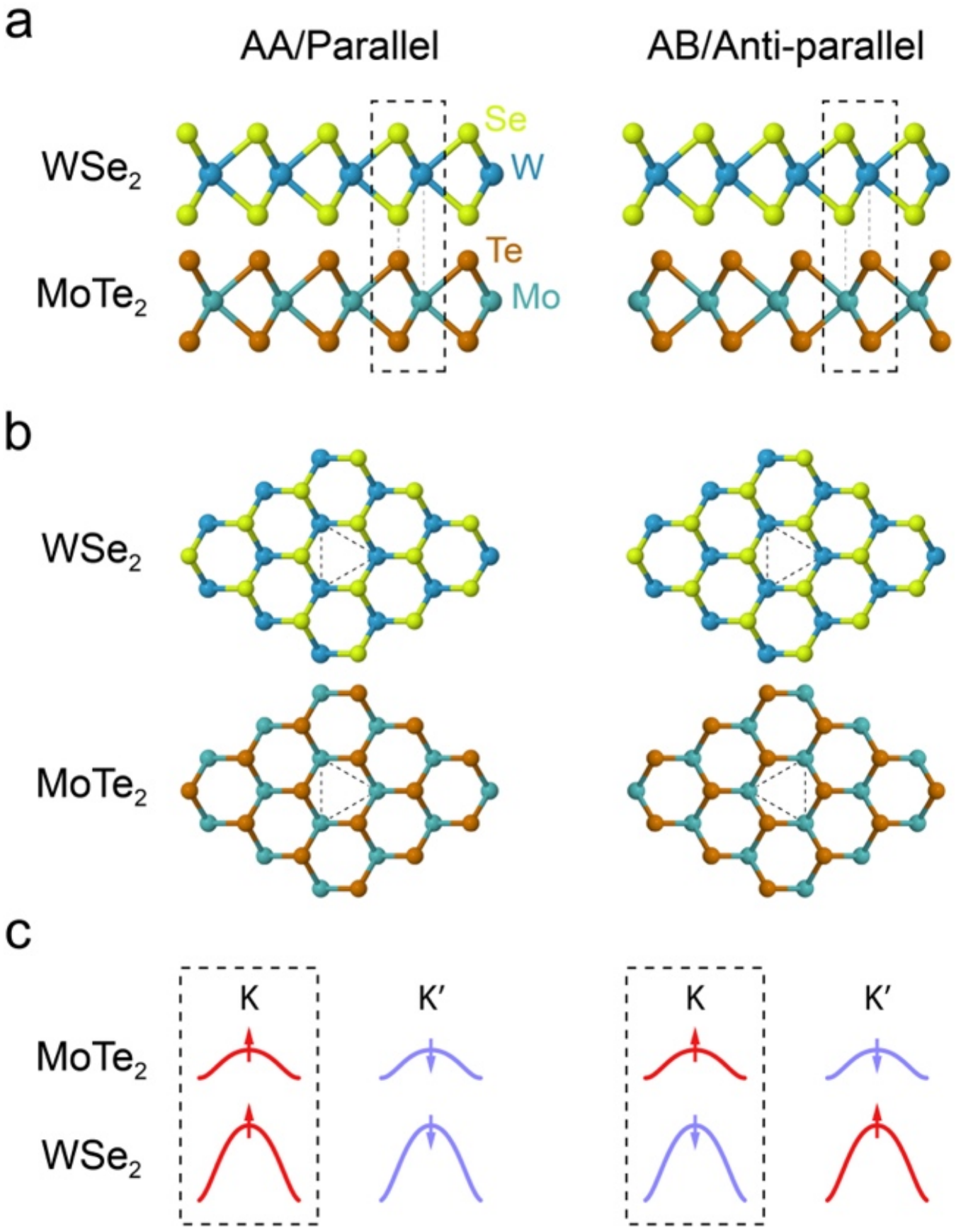

**Figure 1. Stacking order and moiré bands of angle-aligned $MoTe_2$/$WSe_2$ bilayers. a,** Side view. **b,** Top view. The atoms are shown in blue (W), green (Se), cyan (Mo) and brown (Te). **c,** Spin-resolved topmost moiré valence bands from $MoTe_2$ and $WSe_2$. Left: AA-stacking with two layers parallel to each other; right: AB-stacking with two layers antiparallel to each other. In each valley (K/K'), electron spins (arrows) in the two layers are parallel (AA) or antiparallel (AB) due to spin-valley locking in monolayer TMDs. Interlayer hybridization is spin allowed for AA-stacking and spin forbidden for AB-stacking.

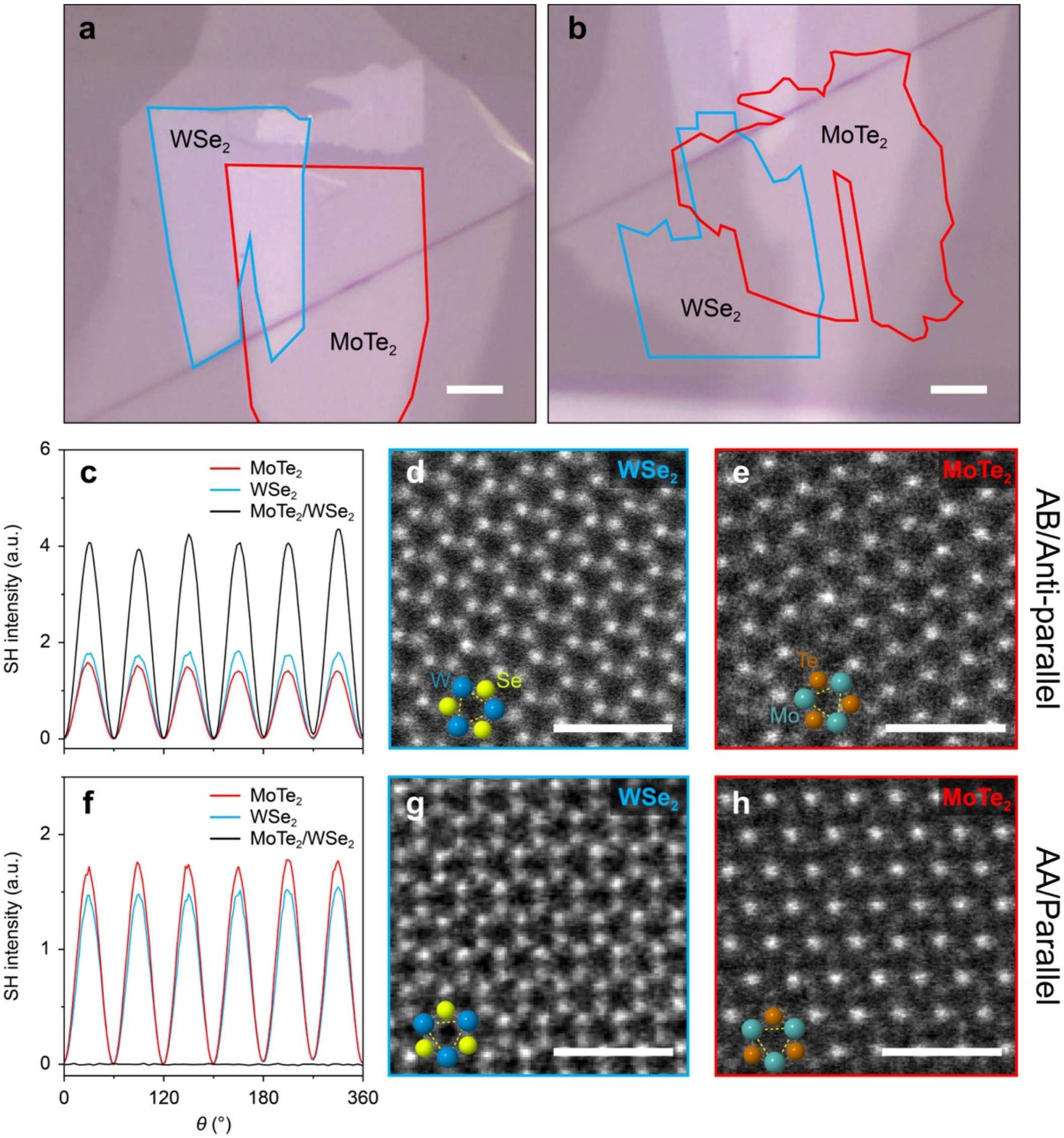

**Figure 2. Stacking order determination. a,b,** Optical image of angle-aligned $MoTe_2/WSe_2$ sample 1 (**a**) and sample 2 (**b**) on PC stamps. The $WSe_2$ and $MoTe_2$ layers are outlined in blue and red, respectively. Scale bars are 5 µm. The $WSe_2$ ($MoTe_2$) layers were cut from the same flake. A 180° rotation was introduced in the two $MoTe_2$ layers. SHG with excitation at 800 nm and STEM measurements were performed on both the isolated monolayer regions and the bilayer region (Methods). **c-e**, Characterization of sample 1. The SH intensity (**c**) exhibits a six-fold rotational symmetry as a function of the excitation polarization angle. The bilayer response is 2-3 times of that of each monolayer. ADF-STEM images from the $WSe_2$ (**d**) and $MoTe_2$ (**e**) monolayer regions identify the atoms (colored balls) and the metal sublattices (marked by yellow dashed lines). Scale bars are 1 nm. **f-h**, Characterization for sample 2. The SH intensity for the bilayer nearly vanishes.

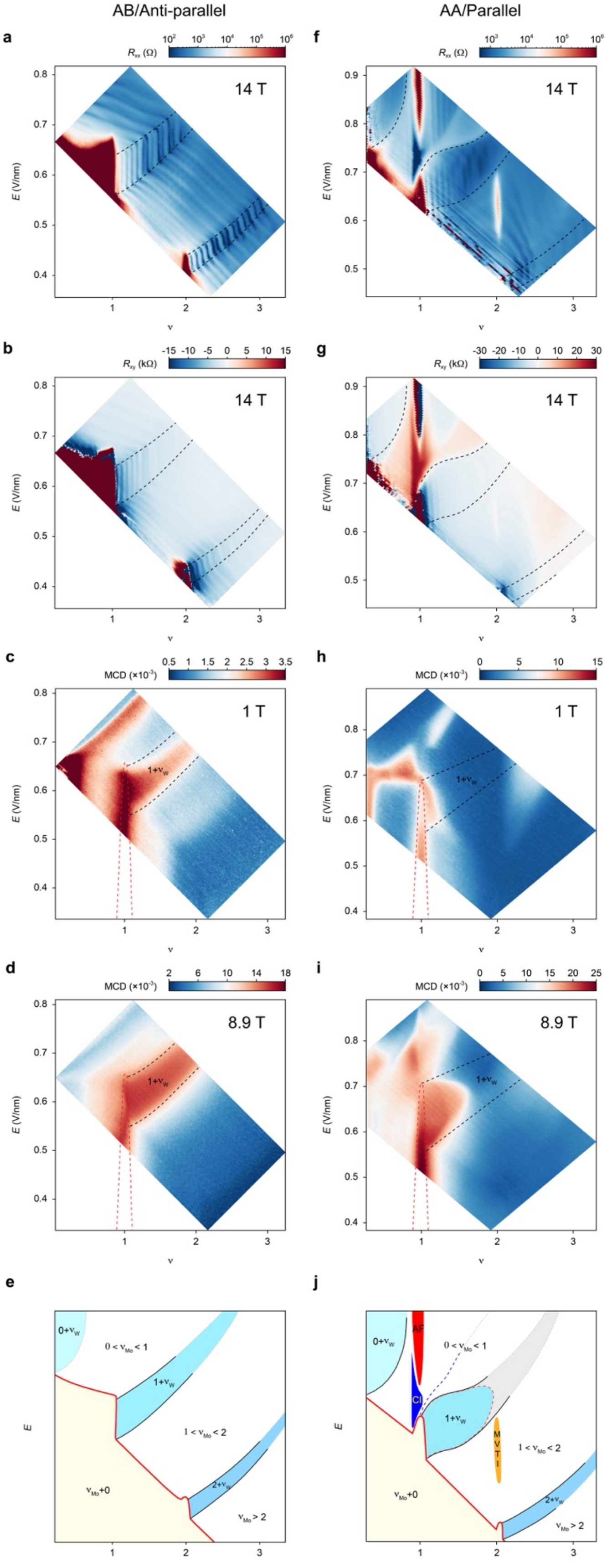

AB/Anti-parallel
AA/Parallel
a
b
c
d
e
f
g
h
i
j
14 T
1 T
8.9 T
$R_{xx}$ (Ω)
$R_{xy}$ (kΩ)
MCD (×10$^{-3}$)
E (V/nm)
ν
E
$1+\nu_W$
$0+\nu_W$
$2+\nu_W$
$0 < \nu_{Mo} < 1$
$1 < \nu_{Mo} < 2$
$\nu_{Mo} > 2$
$\nu_{Mo}+0$
AF
CI
MVTI

**Figure 3. Electronic phase diagrams. a,b,** Longitudinal resistance $R_{xx}$ (**a**) and Hall resistance $R_{xy}$ (**b**) as a function of $\nu$ and $E$ at $T = 1.6$ K and $B = 14$ T . Region $(1 + \nu_W)$ and $(2 + \nu_W)$ (enclosed by dashed lines) exhibit vertical stripes. **c,d,** MCD as a function of $\nu$ and $E$ at $T = 1.6$ K and $B = 1$ T (**c**) and 8.9 T (**d**). The Mott insulating state at $\nu_{Mo} = 1$ and region $(1 + \nu_W)$ show a large MCD response and are marked by red and black dashed lines, respectively. **e,** Schematic phase diagram. Regions are labeled by filling factors in the $WSe_2$ and $MoTe_2$ layers. The sample in **a-e** is an AB-stacked $MoTe_2/WSe_2$ bilayer. **f-j,** Same for an AA-stacked $MoTe_2/WSe_2$ bilayer. New states emerge in the phase diagram (**j**), including a Chern insulator (CI)[29] and an antiferromagnetic (AF) insulator[68] at $\nu = 1$ and a mixed-valence topological insulator (MVTI)[22] at $\nu = 2$. The blue dashed line in region $(0 < \nu_{Mo} < 1)$ separates two Fermi surfaces: one with $R_{xy} > 0$ and another with $R_{xy} < 0$ and vertical SdH stripes; the latter has a small Fermi surface with doping counted from $\nu = 1$. The red dashed line within region $(1 + \nu_W)$ separates a light (blue-shaded) and heavy (grey-shaded) Fermi liquid. The former shows strong MCD.

**Table 1.** ADF-STEM intensity scaling with atomic numbers in $MoTe_2/WSe_2$ bilayers. The high-angle annular dark-field (HAADF) contrast is approximately proportional to $Z^{1.7}$ (Ref.[49]). The contributions from overlapping chalcogen atoms are summed (a.u.: arbitrary unit).

| **Element** | **Atomic number *Z*** | **HAADF contrast ~ $Z^{1.7}$** | **Total contrast (a.u.)** |
|---|---|---|---|
| Mo | 42 | 575 | 575 |
| W | 74 | 1505 | 1505 |
| Se | 34 | 401 | 802 ($Se_2$) |
| Te | 52 | 826 | 1652 ($Te_2$) |

**Supplementary figures**

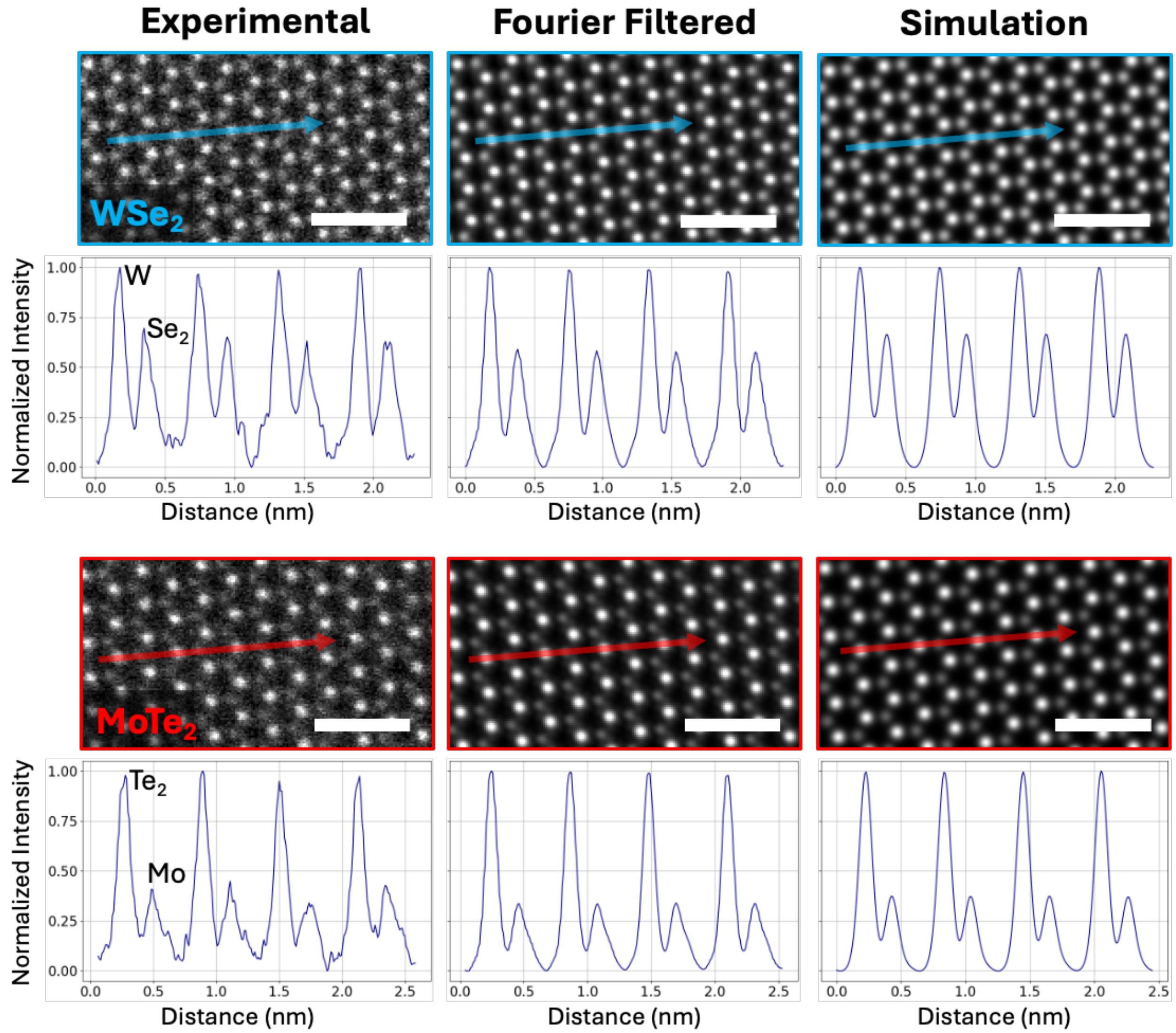

**Supplementary Figure 1. Atomic-resolution imaging and intensity profile analysis.** Top panels display experimental (raw), Fourier-filtered, and simulated ADF-STEM images for $WSe_2$ (blue) and $MoTe_2$ (red) monolayers. Corresponding normalized intensity line profiles (bottom) are extracted along the indicated colored arrows. Accurate atomic species identification is confirmed by the excellent agreement between experimental, Fourier filtered, and the simulated images. All scale bars are 1 nm.

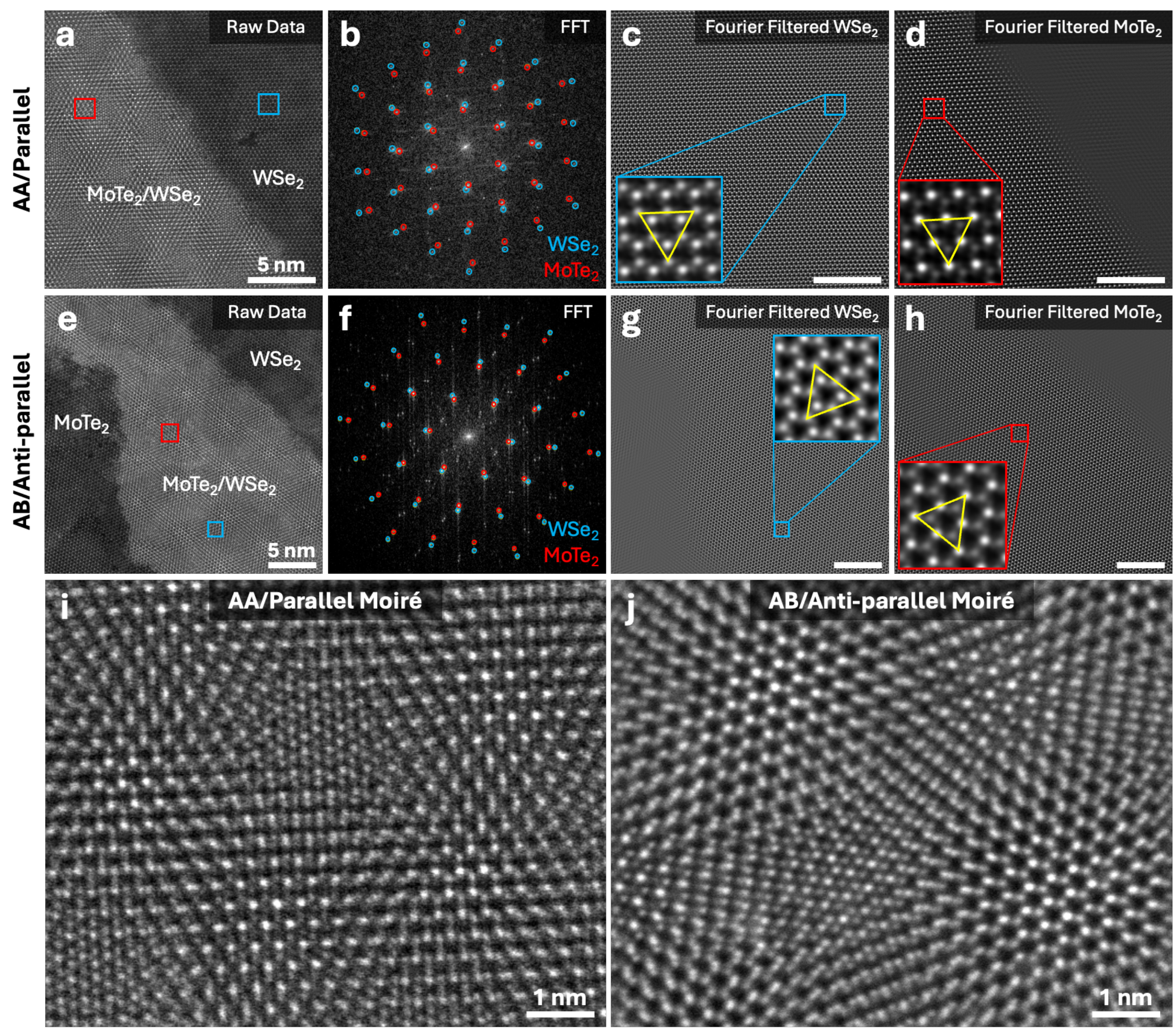

**Supplementary Figure 2. Determination of stacking order of $MoTe_2$/$WSe_2$ bilayers via ADF-STEM and Fourier filtering. a–d**, AA/parallel stacking. Raw ADF-STEM image (**a**) showing the bilayer region and isolated $WSe_2$ region. Fast Fourier transform (FFT) of the bilayer region (**b**), where blue and red circles denote the reciprocal lattice points for $WSe_2$ and $MoTe_2$, respectively. Fourier-filtered images of the individual $WSe_2$ (**c**) and $MoTe_2$ (**d**) lattices. Insets show magnified views with yellow triangles indicating the orientation of the metal-chalcogen sublattices. **e–h**, AB/antiparallel stacking. Scale bars in **a-h** are 5 nm. **i,j,** Moiré pattern of AA-stacked (**i**) and AB-stacked (**j**) bilayers. Scale bars in **i,j** are 1 nm. Determining stacking order via moiré analysis is difficult hence isolated monolayer regions are investigated.

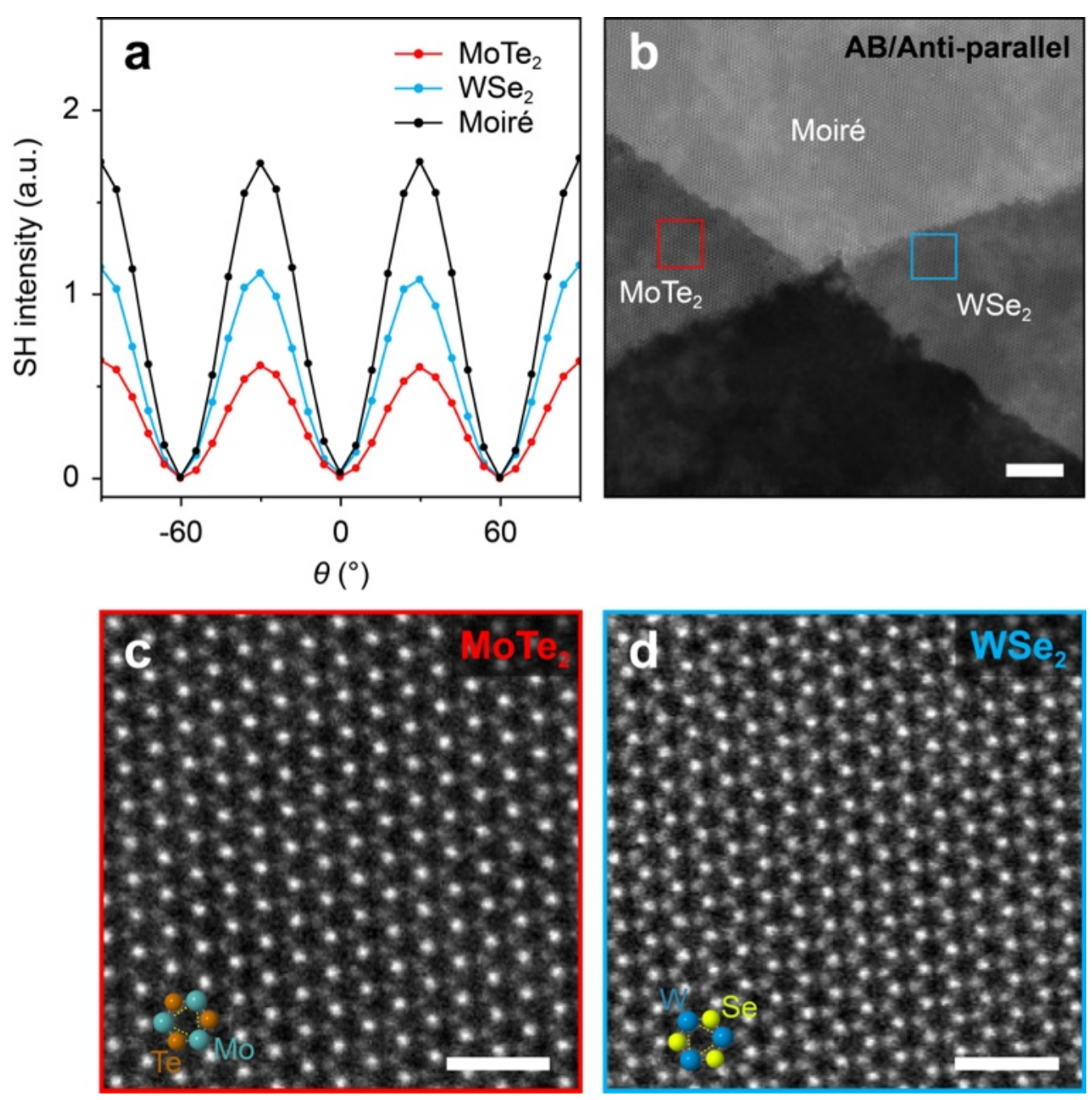

**Supplementary Figure 3. SHG and ADF-STEM characterization on another AB-stacked (antiparallel) sample. a,** SH intensity as a function of the polarization angle for excitation at 800 nm from the bilayer (moiré) region and two isolated monolayer regions. The bilayer region exhibits a significantly enhanced SH response. **b,** Large field-of-view ADF-STEM image covering both the monolayer and the bilayer regions. **c,d,** Zoom-in views of monolayer $MoTe_2$ (**c**) and $WSe_2$ (**d**).

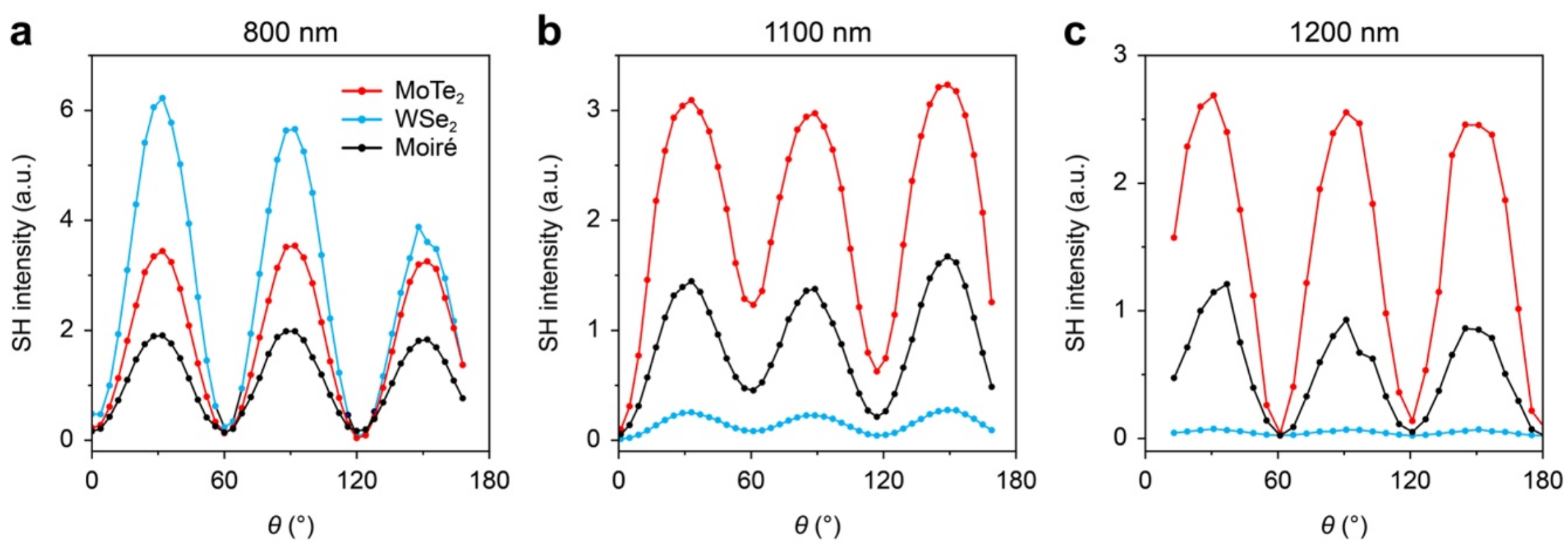

**Supplementary Figure 4. Wavelength dependence of the optical SHG. a-c,** SH intensity as a function of the excitation polarization angle (Methods) obtained from an AA-stacked $MoTe_2$/$WSe_2$ bilayer on a Si substrate. The SHG from the bilayer (moiré) region and the isolated monolayer regions depends on the excitation wavelength. It is also sensitive to sample quality and doping density. Destructive interference of the SH radiation in the bilayer is observed for all three excitation wavelengths (800 nm (**a**), 1100 nm (**b**), and 1200 nm (**c**)), with most complete cancelation at 800 nm. The output of an OPO (optical parametric oscillator, Coherent, Chameleon) operating at an 80 MHz repetition rate was used as the excitation source.

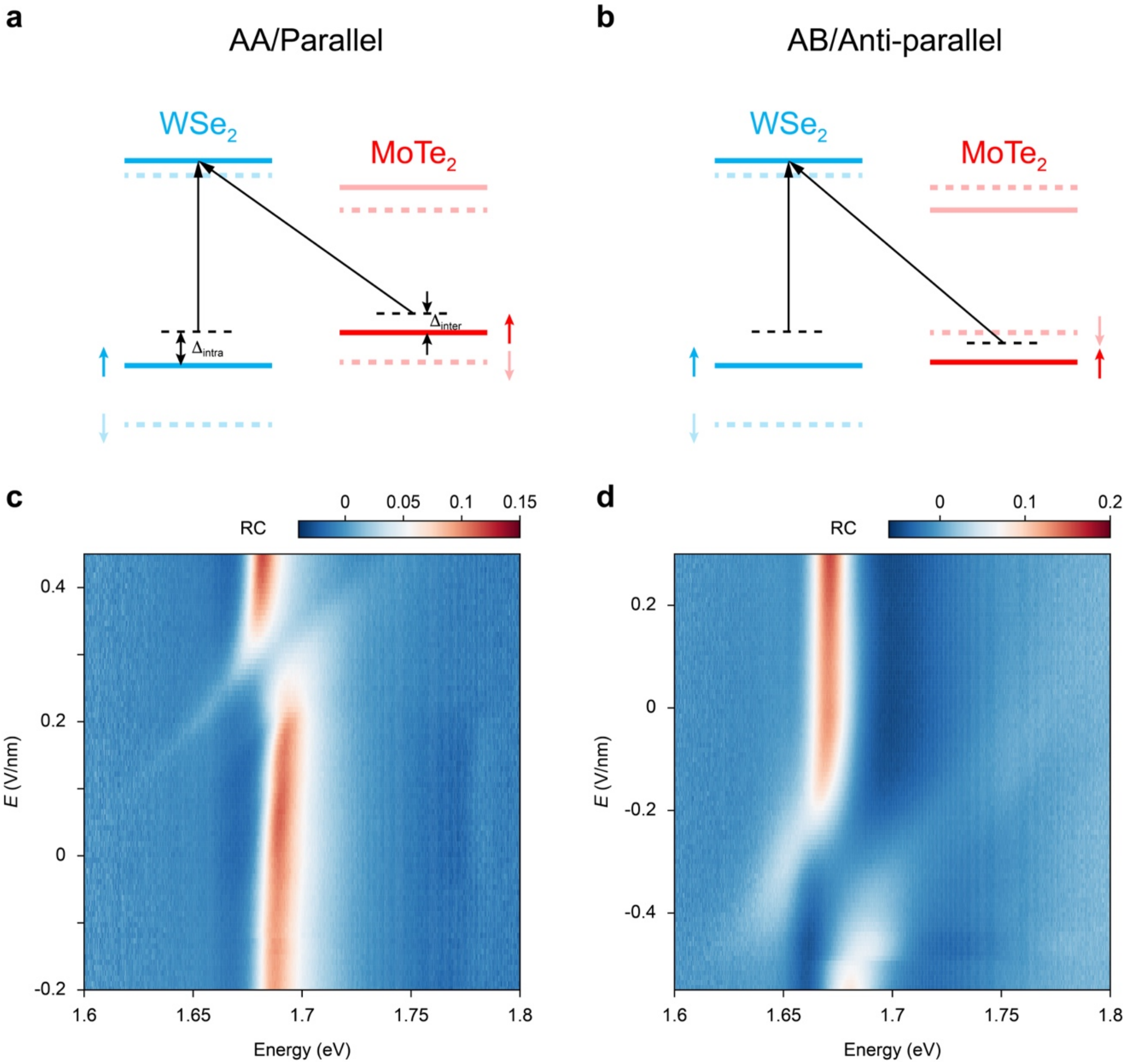

**Supplementary Figure 5. Stacking-order dependent exciton hybridization. a,b,** Spin-resolved band alignment for AA-stacked (**a**) and AB-stacked (**b**) $MoTe_2/WSe_2$ under zero external electric field. The conduction and valence bands of $WSe_2$ (Blue) and $MoTe_2$ (red) from the K valley are illustrated. The spin-up (solid lines) and spin-down (dashed lines) bands are split by the Ising spin-orbit coupling. Black arrows denote intra- and inter-layer exciton excitations in the $WSe_2$ layer. Only spin-conserved optical transitions are allowed; the optically inactive bands are shown in faded colors. $\Delta_{\mathrm{intra}}$ and $\Delta_{\mathrm{inter}}$ denote the intralayer and interlayer exciton binding energies, respectively. See Methods for details. **c,d,** Reflection contrast spectra at $\nu = 1$ as a function of $E$ for AA-stacked (**c**) and AB-stacked (**d**) $MoTe_2/WSe_2$ bilayers.

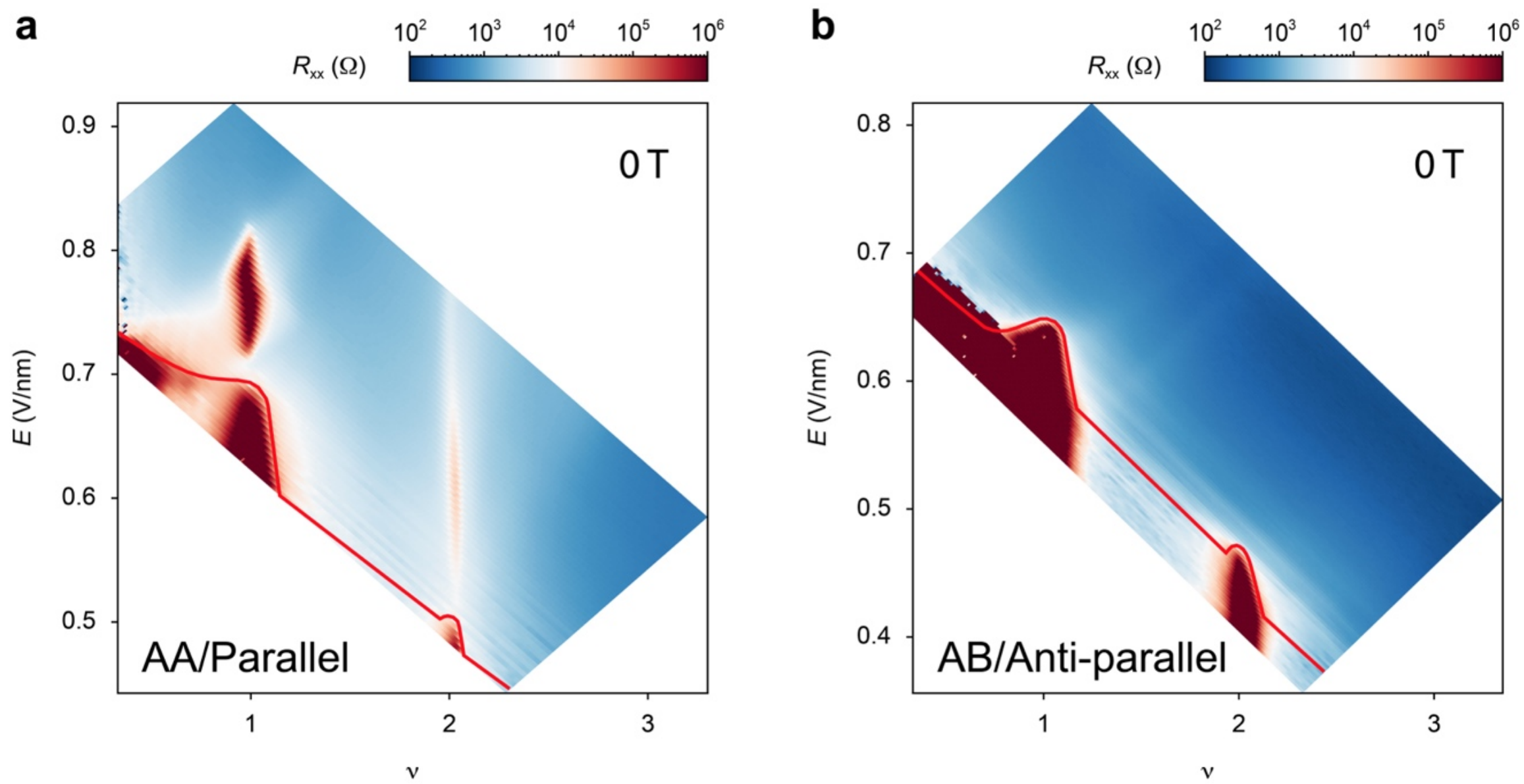

**Supplementary Figure 6. Comparison of zero-field electrical transport results. a,b,** Longitudinal resistance $R_{\mathrm{xx}}$ as a function of $\nu$ and $E$ at $T = 1.5$ K and $B = 0$ T for AA-stacked (**a**) and AB-stacked (**b**) $MoTe_2/WSe_2$. The red curve separates the $\nu_{\mathrm{W}} = 0$ and $\nu_{\mathrm{W}} > 0$ regions. Additional insulating states at $\nu = 1$ and $\nu = 2$ in regions above the red curve are observed only in the AA-stacked sample, consistent with its substantially stronger interlayer hybridization.

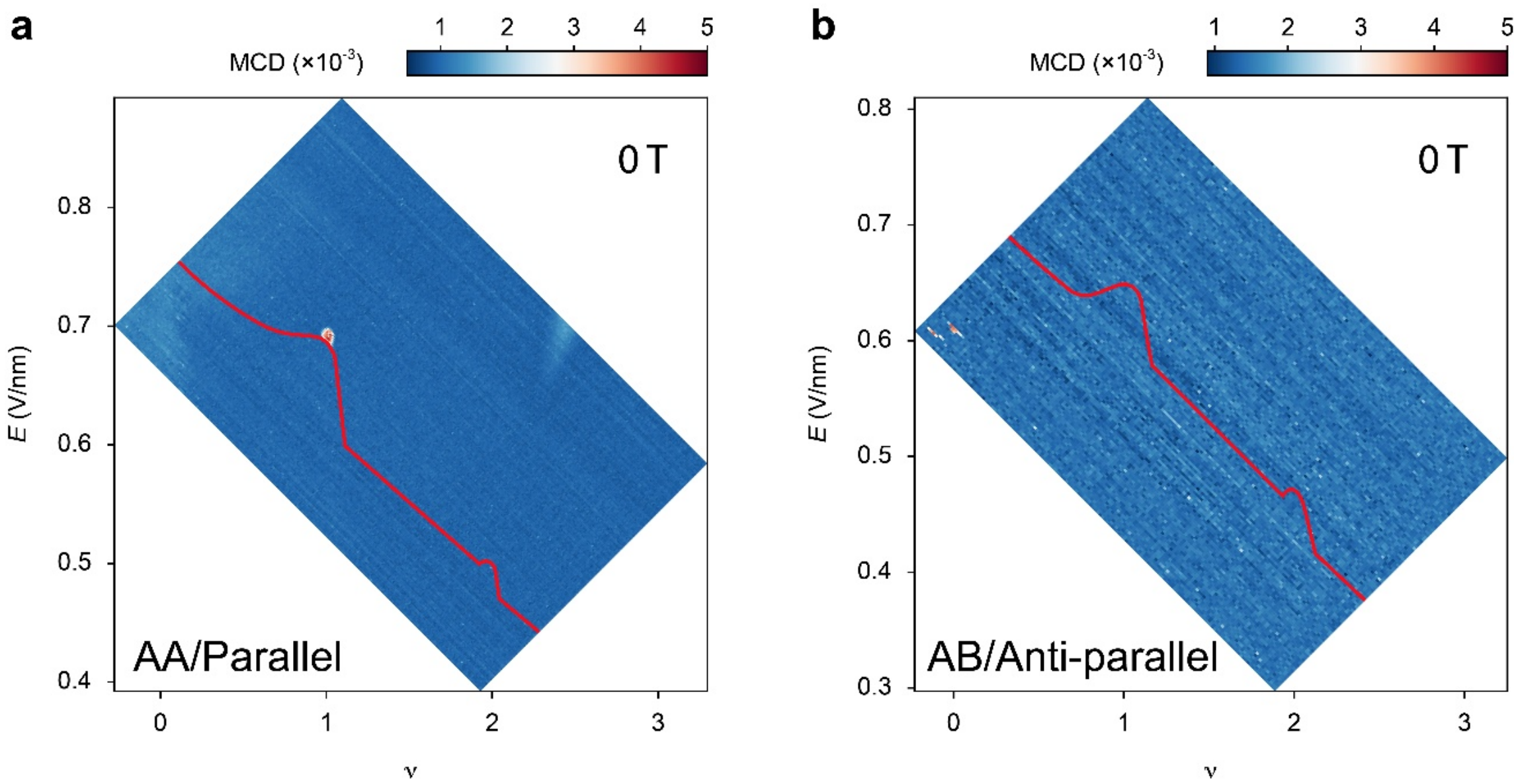

**Supplementary Figure 7. Comparison of zero-field MCD results. a,b,** MCD as a function of $\nu$ and $E$ at $T = 1.6$ K and $B = 0$ T for AA-stacked (**a**) and AB-stacked (**b**) $MoTe_2/WSe_2$. The red curve separates the $\nu_W = 0$ and $\nu_W > 0$ regions. Spontaneous MCD is observed only for the Chern insulator at $\nu = 1$ in the AA-stacked sample.